# Blockchain Technology as a Regulatory Technology
## From Code is Law to Law is Code

Primavera De Filippi[1] & Samer Hassan[2]

[1]CERSA/CNRS & Berkman Center for Internet and Society, Harvard University
[2]Universidad Complutense de Madrid & Berkman Center for Internet and Society, Harvard University

**Abstract:**
"Code is law" refers to the idea that, with the advent of digital technology, code has progressively established itself as the predominant way to regulate the behavior of Internet users. Yet, while computer code can enforce rules more efficiently than legal code, it also comes with a series of limitations, mostly because it is difficult to transpose the ambiguity and flexibility of legal rules into a formalized language which can be interpreted by a machine. With the advent of blockchain technology and associated smart contracts, code is assuming an even stronger role in regulating people's interactions over the Internet, as many contractual transactions get transposed into smart contract code. In this paper, we describe the shift from the traditional notion of "code is law" (*i.e.* code having the effect of law) to the new conception of "law is code" (*i.e.* law being defined as code).

**Introduction**
There are various ways in which law and technology can influence each other. The two interact through a complex system of dependencies and interdependencies, as both contribute (to a greater or lesser extent) to regulate the behavior of individuals. With the advent of modern information and communication technology, the relationship between the two has significantly evolved, as the latter is increasingly used as a complement or a supplement to the former. Lawyers, judges and policy makers are increasingly surrounded by digital information and software tools, which they use in their daily routine. While these tools can be used to support their activities, technological innovation also raises a variety of challenges, which the legal profession will eventually need to address. Specifically, it is possible to identify four distinct phases, in the late 20th and early 21st century, that represent the evolving relationship between law and technology.

The **first phase** involves the process of **digitizing information** —turning paper and ink into computer readable information. That phase is now well under way: copies of cases, statutes, and regulations have been available on-line for decades in large databases (Berring, 1986), accessible at first for a fee, and now mostly for free.

The **second phase** consists in **bringing automation to decision-making processes**. Most of the legal informatics research to date has focused on translating legal provisions into computer code. Both policy makers and judges are increasingly relying on computer applications (e.g. expert systems, as in Waterman et al, 1986) to retrieve legal provisions or case law, analyse or compare them, so as to build a proper argumentation and ideally come up with better decisions. This is a difficult task for many different reasons, including the ambiguity of human language and the need for legal norms to be flexible and fact dependent. Despite these challenges, governmental institutions and businesses worldwide increasingly rely on rule-based representations of specific knowledge domains (such as health care and tax or financial regulations) for automated or semi-automated decision-making (see e.g. specific software tools for taxation, accounting and credit-score assessment).

The **third phase** involves the **incorporation of legal rules into code** on the one hand, and the **emergence of regulation by code** on the other. With the widespread deployment of the global Internet network, new forms of regulation have emerged which increasingly rely on *soft law* (*i.e.* contractual agreements and technical rules) to regulate behaviors. As more and more of our interactions are governed by software, we increasingly rely on technology not only as an aid in decision-making but also as a means to directly enforce rules. Software thus ends up stipulating what can or cannot be done in a specific online setting more frequently than the applicable law, and frequently, much more effectively. This is what Joel Reidenberg has coined *Lex informatica* (1992) —a concept which has been subsequently been popularized as "*Code is law*" by Lawrence Lessig (1999).

Regardless of the terminology used, the core characteristics of this new type of law is that it relies on code in order to define the rules that people need to abide by. On the Internet, regulation is mostly done by private means (*e.g.* by the device or software designers) in an environment which, because of its transnationality, seemed (at least initially) to exist beyond the jurisdiction of the nation states.

An emblematic example of that are Digital Rights Management (DRM) schemes, transposing the provisions of copyright law into technological measures of protection (Rosenblatt et al, 2001), and thus restricting the usage of copyrighted works (e.g. by limiting the number of possible copies of a digital song that can be made). The advantage of this form of *regulation by code* is that, instead of relying on *ex-post* enforcement by third-parties (*i.e.* courts and police), rules are enforced *ex-ante*, making it very difficult for people to breach them in the first place. Besides, as opposed to traditional legal rules, which are inherently flexible and ambiguous, technical rules are highly formalized and leave little to no room for ambiguity, thereby eliminating the need for judicial arbitration.

More recently, a new technology has emerged which might change the way we think about law. This technology is the blockchain, a decentralized, secure and incorruptible database (or public ledger) that constitutes the foundational tool for peer-to-peer value creation and trustless transactions. Introduced in 2009 with the Bitcoin network —as the underlying infrastructure for a decentralized payment system— the technology has rapidly evolved to acquire a life of its own.

Today, the blockchain is used in many other kinds of applications, from financial applications to machine-to-machine communication, decentralized organizations and peer-to-peer collaboration. As a trustless technology, the blockchain eliminates the need for trust between parties, enabling the coordination of a large number of individuals that do not know (and therefore do not necessarily trust) each other.

At the very end of the spectrum, the most recent blockchains have introduced the ability for people to upload small snippets of code (so-called *smart contracts*) directly onto the blockchain, for them to be executed in a decentralized manner by every node of the network. These rules are automatically enforced by the underlying technology (the blockchain), even if they do not reflect any underlying legal or contractual provision.

This is what brings us to the **fourth phase** —which is just beginning— involving a new approach to regulation, the ***code-ification* of law**, which entails an increasing reliance on code not only to enforce legal rules, but also to draft and elaborate these rules. As a result of these technological advances, the lines between what constitutes a legal or technical rule becomes more blurred, since smart contracts can be used as both a support and as a replacement to legal contracts.

Indeed, even though the majority of smart contracts are not directly associated with an actual legal contract, depending on how they have been entered into, they may or may not give rise to an actual contractual relationship in the traditional meaning of the word. Yet, from a purely technological standpoint, smart contracts can be used to emulate, or at least simulate the function of legal contracts through technology, thereby effectively *turning law into code.*

The focus of this paper is on the last two phases. Section I provides an overview of the third phase, analyzing the specificities of code (I.A), the various benefits and drawbacks of regulation by code (I.B) and the ways in which law has, thus far, attempted to regulate code (I.C). Section II investigates the fourth phase, introducing the blockchain paradigm (II.A), along with the distinctive features of blockchain code (II.B) and the extent to which it can be regarded as a regulatory technology (II.C).

## I.     Code is Law

The role of technological artifacts as an enforcement tool existed long before the advent of modern information technologies. Technological artifacts are not neutral, but inherently political (Mowshowitz, 1984): even if they are often defined as general-purpose technologies, their design will ultimately dictate the type of actions that they might enable or prevent. According to Winner (1980), political choices are —either intentionally or unintentionally— embedded into the design of a technology, and these different designs will, in turn, have important societal implications, insofar as they might support certain political structures or facilitate certain actions and behaviors over others. This can be observed in the context of urban planning —*e.g.* the roads of many cities have been constructed in such a way as to conceal the view of slums from the city center, and public benches in the poorest neighborhoods are often designed to prevent people from sleeping on them (Smith, 1996). But the same is true in many other sectors of society. For instance, in the context of energy production and distribution, the deployment of nuclear plants is likely to lead to a much more centralized system than the reliance on solar panels. The former involve centralized distribution of energy, as they require strong centralized control and high security measures (e.g. an army to avoid terrorist attacks); whereas the latter promote a much more decentralized distribution of energy, empowering individuals with greater authority and autonomy concerning their own energy production and consumption (Winner, 1980).

According to Lessig (1999), four different forces exist, which all contribute —to a greater or lesser extent— to shaping individuals' actions, in ways that often remains outside of the control of any given individual alone.
- *Law* creates artificial constraints limiting individual actions through legal rules and regulations (e.g. by making it forbidden for people to steal, and punishing those who infringe these rules).
- *Social Norms* regulate cultural behaviors through peer pressure (e.g. by making it socially unacceptable for people to speak loudly in a professional meeting).
- The *Market* encourages or discourages specific behaviors through the mechanism of supply and demand (e.g. by setting prices for specific goods or services).
- *Architecture* (defined by Lessig as "features of the world, whether made, or found") imposes a series of constraints by limiting the type of actions that an individual can do (*e.g.* biology, geography, technology are all, to some extent, constraining people's actions).

*Architecture* does not, however, sufficiently account for the distinction that exists between *Nature* and *Technical artefacts* (including the urban setting). Both impose constraints by limiting the type of actions that an individual can do (e.g. water limiting one's ability to travel through it, or benches whose design is precluding one's ability to sleep), but the former is a given (*by nature*) whereas the latter is the result of a deliberate choice (*by people*).

The widespread deployment of information technologies and the global Internet network have created a new environment for human expression, whose rules are mediated mostly (if not solely) by software code. Just like any other technological artifact, this code might reflect political interests, and its technological design might have important implications over the online experience of many individuals (Winner, 1980).

As clearly stated by Lessig's "Code is Law" (1999), code is ultimately the architecture of the Internet, and —as such— is capable of constraining an individual's actions via technological means. The architectural implementation on online platforms ultimately depend on the specific choices of the platform designers, seeking to promote or prevent a certain type of actions. But how far can one predict, or even orchestrate the effects that a particular technology might have? An important difference between the physical and digital world is that, even though a single individual cannot influence forces like *Law*, *Social Norms*, *Markets*, and *Nature*, individuals are increasingly able to create and manipulate code (either by themselves, or by getting others to do it). Of course, while the design of any technological artefact can be carefully enrobed with political intentions, the correlation between the technical design and political implications of a technology are not always evident. Although a technological infrastructure might be designed to promote or prevent certain types of behaviors, it is not always guaranteed to have the desired effects. Indeed, technological artefacts are constantly used and reused for different purposes depending on the contingencies. The implications of a particular technology cannot therefore be fully understood without accounting for the social and historical context in which the technology operates. Rather than the technological design, it is, ultimately, the way a technology is adopted by a particular group of individuals that will determine its social and political impact (Jeorges, 1999).

Whether the consequences are intentional or not, the digital environment opens up the doors to a new form of regulation —by private actors— which might try to impose their own values by embedding them into a technological artefact. Depending on whether, and how, these technologies will subsequently be adopted by people (Woolgar and Cooper, 1999), they could potentially have a significant impact on a very large number of individuals.

**I. The Specificity of Code**

Code, and specifically Internet code, possesses specific features that make it inherently different from other forms of regulation. Building software applications is quite different from building and engineering hardware devices (or any other physical good for that matter). Both processes follow similar logical patterns and rely on similar methodologies. Yet, the fundamental difference between the two is that —as opposed to physical artefacts, whose production requires raw materials and (often expensive) production facilities— code can be produced with just a computer and can be distributed via any kind of storage device or network connection. Hence, the **barriers to entry** for software builders are significantly **lower** than in many other contexts (Campbell-Kelly 1995) —as can be witnessed by the exponential growth of software applications in the past few decades.

Code displays many other interesting characteristics. As opposed to the physical world, where the cost of reproduction is generally quite high, in the digital realm, the **cost of reproduction** is virtually **zero**[1]—or "zero marginal cost" as per Rifkin (2014). This means that software code may be easily copied, modified and spread around the globe, with a significantly high speed. Moreover, given that software code is in a digital format, anyone can replicate the code and modify it at will in order to create an alternatively version of it (i.e. a fork). The digital nature of code thereby ensures a higher degree of **adaptability** and **malleability**.

With digital technology, the cost of distributing information is also close to zero. As long as two or more devices are connected (forming a network) through a physical cable or radio signals, they may transmit information to one another at zero cost. In the context of a transnational network like the Internet, any piece of code can quickly be reproduced (and adapted) in multiple areas of the world, regardless of national boundaries or other jurisdictional matters. Thus, it becomes difficult for a State to prevent the exportation or importation of code. Indeed, several attempts at restraining the dissemination of a sensible software code have repeatedly failed in the past—see *e.g.* the RSA cryptographic algorithm (Evans 1993) or the DeCSS decryption of DVD DRM (Eschenfelder and Desai 2004).

Finally, as opposed to traditional legal rules which can only be enforced *ex-post* (*i.e.* after the fact), regulation by code can impose limitations on individual actions in a way that can be enforced *ex-ante* (*i.e.* the code is preventing people from breaching the technical rules, even before they act). Savvy individuals and technical experts may be capable of finding workarounds, but the majority of the people — those without specific knowledge or resources— have no choice but to comply with the rules of the code. This is very different from regulation by law, which empowers people with the ability to decide whether or not to infringe the rules, and rely on using courts and police to enforce the rules only after they have been broken (Lessig 1997).

**I.B Regulation by Code**

Law and technology enjoy a complicated, and to a large extent interconnected, relationship. On the one hand, the State is struggling to exercise its sovereignty over the Internet, by regulating code in order to (indirectly) regulate individual users. On the other hand, code is increasingly employed in a wide variety of sectors[2] to regulate behaviors — either jointly with, or in addition to, existing laws.

---

[1] Of course, this is without accounting for the cost of the infrastructure, in the case of local copies being a computer/device, or in the case of Internet copying, the cost of the global communication infrastructure that sustains the network.

[2] Neyland (2006) provides a detailed list of socio-technical systems used to regulate individuals, these include: "Systems for the regulation of identification (such as biometrics), movement (such as traffic management) and non-movement (such as airport security), disposal (such as waste management), saving (such as banking regulations), spending (such as point of sale machines) and retrieval (such as illegal download tracking) form just some of the many developments in this area."

The most emblematic example of code being deployed to incorporate and enforce existing legal provisions can be observed in the context of copyright law. Copyright law was originally intended to protect authors and creators from free-riding, by granting them a set of moral and exclusive rights (as enshrined in the Berne Convention) with a view to enable them to make a living by licensing or transferring these rights (Ricketson, 1987). However, the advent of the Internet substantially changed the landscape in which those agreements were signed. First, the enforcement of these exclusive rights became extremely difficult in the digital world, mostly due to the ease of reproduction and distribution of digital works. Second, digital technologies spurred the emergence of the *Free Culture movement* (Fuster Morell, 2012) advocating for the right to freely reproduce, distribute and remix creative works — in clear conflict with the commercial approach of many right holders, eager to preserve the "all rights reserved" regime of copyright law (Aigrain, 2012). To protect the economic interests of these right holders, many content providers started using Digital Rights Management (DRM) systems and Technological Protection Measures (TPM), with a view to restrict the possible uses of digital content by end-users, through a series of access control or copy control mechanisms (Samuelson, 2003).

The advantage of DRM systems is that they facilitate the process of copyright enforcement by enabling right holders to dictate the manner in which users can access or consume a work by technological means. This include, for instance, technical protections designed to preclude the reproduction of MP3 files, the copying of DVDs, the modification of PDF documents, or the remix of multimedia files. These technologies have been embraced by the traditional content industry (e.g. Disney, Time Warner) but also by new industrial actors whose successful business models heavily depend on DRM (e.g. Netflix or Valve's Steam).

However, this advantage comes at a cost. In fact, many legal provisions that cannot be easily incorporated into these technological systems (such as copyright exemptions and fair uses) are generally ignored by the system, often at the detriment of end-users. Thus, in addition to protecting against copyright infringement, many DRM systems also prevent users from legitimately accessing or reproducing copies of a work, since the code rarely differentiates among the different types of users (e.g. end-users, libraries, corporations) and uses (e.g. educational uses, non-commercial, parody, etc.). Thus, we observe a discrepancy between what these technological artefacts are meant to achieve (*i.e.* preventing people from engaging in copyright infringement) and the actual effects they have on society (Woolgar and Cooper, 1999). Whether or not this was deliberately intended by those who designed the technology, the effect of their technological design is such as to significantly impair the right for people to access and distribute information online.

Of course, the code of a DRM system can also be circumvented by code. A wide variety of workarounds have been implemented to circumvent DRM restrictions (Samuelson, 2003), with several technical solutions involving e.g. cracks, rooting procedures, decryption techniques (see e.g. the memorable DeCSS algorithm that decrypted the DVD codec). These are usually released as open source software, and be made available to anyone with the necessary knowledge or resources to use them. In order to avoid that from happening, anti-circumvention laws were enacted in many countries —following the incorporation of these provisions within the World Intellectual Property Organization's Copyright Treaty of 1996—to prohibit people from bypassing technological protection measures applied to protected digital content in ways that were not authorized by the relevant rights-holders (Besek 2003). As a result, code is used to reinforce the law, as a more efficient means to address the complexity of copyright enforcement in the digital realm, and law has been used as a tool to strengthen the code, to ensure that it could not be circumvented or tampered with.

In addition to supporting or complementing the law, code can also be deployed as a way to avoid or bypass the law. The popular Napster case is a good example of that (Ku, 2002). Launched in 1999, Napster presented itself as a web service providing users with the capability of sharing musical files with each other. However, the company faced several legal difficulties related to copyright infringement issues, and was rapidly forced to shut down the service. In order to avoid a similar outcome, decentralized protocols for peer-to-peer (P2P) file sharing (such as BitTorrent) have been subsequently implemented, to avoid the need for a central point of failure (or control) which could be legally prosecuted and shut down (Pouwelse et al, 2005). Up until now, all legal attempts at shutting down BitTorrent have failed — an illustration of how software code can successfully be used to circumvent law-originated rules.

Finally, code may introduce new rules which have little or nothing to do with existing laws. For instance, many P2P file sharing protocols embed in their code the requirement for users to share content before they can download more content, thereby enforcing some form of cooperation among users. But the effect of code in shaping online behavior goes much deeper than that. Most relevant in this context is the function of Graphical User Interfaces (GUI), whose design has been extensively studied (in the fields of Human Computer Interaction and Science and Technology Studies) to analyze the social and political implications it engenders (Kannabiran and Petersen, 2010; Patrick and Kenny, 2003). Online service providers frequently rely on code (or algorithms) to affect or influence the behavior of their user-base (Pasquale, 2015). In this regard, Facebook has been often criticized for its obscure and inaccessible privacy settings (Hargittai, 2010), for its Orwellian social environment (where whomever oversteps the blurry lines of the Terms of Use is expelled from the platform), for its oversimplification of human emotions (through the "Like" button), for manipulating user emotions and interactions (Kramer et al., 2014) and for promoting individualistic and narcissistic values, only with a view to maximize its own profits (Lovink 2013). This kind of paralegal regulation can also be achieved via hardware-software integrated systems, such as CCTV surveillance cameras, which modify user behavior sometimes in unexpected ways (Neyland, 2006). In all cases, the code and underlying algorithms may have implicit biases (many times unforeseen) that might result in discrimination, unfairness, surveillance, or questionable associations (Gillespie, 2014; Ziewitz, 2016).

## I. C Law regulating Code

Contrary to what some early proponents have suggested (see *e.g.* the Declaration of Independence for the Cyberspace, in Barlow 1996), the Internet does not exist in a vacuum. Online operators, software developers and device manufacturers incorporated as legal entities in a particular jurisdiction are subject to the laws of that country, whether they want it or not.

Hence, although it is difficult for governments to enforce their laws in a transnational environment like the Internet, it is always possible for States to enact local regulations aimed at regulating software developers or device manufacturers —see *e.g.* the *Clipper chip* backdoor in the U.S.[3]— and online intermediaries —*e.g.* by introducing monitoring requirements[4] and disclosure obligations,[5] or by requiring that they abide to minimum privacy requirements.[6]

The regime of intermediary liability limitations is a broad set of statutory limitations on liability which apply to online operators, such that content transmitted through or stored on their infrastructure does not expose them to criminal or civil liability. However, the enactment of the Digital Millennium Copyright Act in 1998, the European E-Commerce Directive in 2000 and the Information Society Directive in 2001 spurred a trend —which has continued since then— where intermediary liability limitations are increasingly subject to the arbitrary judgement of right holders. For instance, the rules for notice and takedown introduced by the Digital Millennium Copyright Act leave online providers little choice other than to comply with right holders' demand, even if they suspect the takedown order to be erroneous. The SOPA and PIPA proposals brought up in late 2011 and early 2012 went further in this direction by proposing to introduce liability not only for hosted copyright violations, but also for linking to, advertising on, or providing financial services to any site that could host infringing material (Benkler et al., 2015). Finally, the Anti-Counterfeiting Trade Agreement (ACTA)—which has been rejected in the European Union, but not elsewhere[7]— attempted to establish higher standards for the enforcement of intellectual property laws against piracy and counterfeiting, but it did so by giving rights holders greater leeway to hold online operators liable for hosted content that they claim is infringing, without option of judicial recourse (Matthews and Žikovská, 2013).

In the end, both SOPA and PIPA failed to pass due to public backlash. Similarly, the ACTA received strong opposition for trying to turn online operators into a "private police" of the Internet. Yet, these are emblematic of the fact that, by regulating the actors in charge of developing, managing and operating today's online platforms, governments are also capable of —albeit indirectly— regulating the users interacting with these platforms.

---

[3] In the '90s, the U.S. government tried to impose the incorporation of the 'Clipper chip' —for which it had a back-door key, into every device manufactured by the U.S. industry.
[4] *C.f.* European Data Retention Directive (2006)
[5] *C.f.* U.S. PATRIOT Act and the Foreign Intelligence Surveillance Act Amendment (FISAA).
[6] *C.f.* European Directive on Data Protection (1995) and the coming Data Protection Regulation.
[7] Signatory countries to the ACTA include Australia, Canada, Japan, South Korea, Morocco, New Zealand, Singapore and the US, none of which has ratified ACTA yet, except for Japan.

## II. Law is Code

The idea that *Code is Law* has now become a popular conception (Wu 2003). Over the years, following the widespread deployment of the Internet network and our growing dependency on digital technologies, there has been a tendency by private actors (and public institutions) to replace current laws and regulations —which can only be enforced *ex-post* through State intervention— by technical regulation —which can be enforced *ex ante* through code.

Yet, the practice of transposing legal rules into technical rules is not an easy task. As opposed to legal rules, written as general rules in a natural language that is inherently ambiguous, technical rules can only be implemented into code, and thus necessarily rely on formal algorithms and mathematical models. Regulation by code is therefore always more specific and less flexible than the legal provisions it purports to implement.

Transposing legal rules into technical rules is, therefore, a delicate process that could have an important impact on the legal system, and which might actually affect the way we think about law. The inherent ambiguity of the legal system —necessary to ensure a proper application of the law on a case-by-case basis— ultimately gives software developers and engineers the power to embed their own interpretation of the law into the technical artefacts that they create.[8]

Hence, while it is true that, in the digital world, code is increasingly assuming (and perhaps even replacing) some of the traditional functions of law, it is also true that, in the last few years (especially since the emergence of blockchain technology and corresponding smart contract transactions) the law is progressively starting to assume the characteristics of code.

Blockchain technology reinforces the tendency to rely on code (rather than on the law) to regulate individual actions and transactions. The blockchain enables a whole new type of regulation by code, which —combined with smart contracts— also promotes a new way of thinking about the law. Indeed, as more and more contractual rules and legal provisions are incorporated into smart contract code, the traditional conception of the law (as a flexible and inherently ambiguous set of rules) might need evolve into something that can better be assimilated into code. As a result of this tendency, both lawyers and legislators could increasingly be tempted to *deliberately* draft legal or contractual rules in a way that is much closer to the way technical rules are drafted. *Code is Law* might therefore lead to *law progressively turning into code.*

---

[8] This is clearly illustrated by the work of Lucas Introna and Niall Hayes (2004) concerning the biases found in plagiarism detection systems. Due to the difficulty for a technical system to recognize plagiarism at the semantic level, most of these automated systems focus on verbatim similarities and ignore the practices of legitimate quotation. In the words of Introna: "The assumption that copying is equal to plagiarism, which is embedded in the algorithm of these systems and in the way they are often used, can lead to very unfair outcomes for some students." [Commentary by Francesca: is this quote coming from the Introna/Hayes reference? If yes why only attributed to Introna? If no, can you insert the correct reference?]

## II. A The Blockchain Paradigm

The blockchain has been first introduced by Bitcoin, a decentralized payment system that operates independently of any government or central bank. A blockchain is a decentralized database (or state machine) that relies on a set of cryptographic primitives to ensure data integrity and authenticity. Data stored in a blockchain cannot be retroactively modified, so that the 'state' of the blockchain can only be updated through consensus (*i.e.* with approval of more than 50% of the network nodes) by adding new data to it. In this sense, the blockchain can be regarded as cryptographically secure append-only database, which operates without the need of any central authority or clearing house.

As opposed to the Bitcoin blockchain, which was specifically designed to operate as a decentralized payment system, modern blockchain architectures (such as Ethereum, in Buterin 2014) introduced additional functionalities, allowing for small snippets of code to be deployed directly on the blockchain and to be executed in a decentralized manner by every node in the network. These are commonly referred as **smart contracts** (SC), in that they enable people to enter into a contractual relationship with other people (or machines) through a simple transaction on the blockchain.

Smart contracts were first described by Nick Szabo in the late 1990s. Szabo (1997) envisioned placing contracts into code so that they could be both "trustless" and self-enforcing, thereby enhancing the efficiency and removing the ambiguity of traditional contractual relationships. Beyond increased speed and efficiency, an important benefit of smart contracts over traditional contracts is the lack of textual ambiguity, as their provisions are written in a formal language that must be understood by a machine.

Smart contracts aim to emulate the logic of contractual clauses. They are computer programs that facilitate the negotiation, verify and enforce the performance of a contract, or that can even obviate the need for an underlying contractual agreement between parties (Szabo, 1997). In fact, smart contracts are able to automatically execute the terms of a specific agreement, providing trustless transactions via integrated enforcement mechanisms.

As such, smart contracts can support the performance of contracts, reducing costs of negotiation, verification and enforcement by turning legal obligations into self-executing transactions. Earlier examples of (non blockchain-based) smart contracts are traditional vending machines; phone locking by telecom providers; DRM systems; cars incorporating automated speed limitations; etc.

When smart contracts are implemented on a blockchain, their execution is not performed on a central server, but is rather distributed amongst the network of nodes. Blockchain-based smart contracts are therefore more sophisticated than traditional means of technological regulation in that they qualify as computer software code which is both autonomous— as it does not depend on any given third party to operate, and independent — as it cannot be controlled by anyone (De Filippi and Wright, 2015).

Smart contracts can interact with both humans and other smart contracts within the same blockchain ecosystem (e.g. Ethereum). In some cases, a complex set of smart contracts is set up in such a way as to make it possible for multiple parties (SCs or humans) to interact with each other. This combination of smart contracts may be regarded as a **distributed autonomous organization** (or DAO) — a self-governed organization[9] controlled only and exclusively by an incorruptible set of rules, implemented under the form of a SC. An individual may decide to transact with the DAO in order to, for instance, get paid in exchange of a service.[10] Thus, a DAO could in practice *hire* people or SCs to perform specific tasks, and could potentially *sell* their own services (or resources) to third-parties. DAOs operate thanks to all network's nodes; they do not rely on any central server and thus cannot be shut down (unless they feature an explicit *kill-switch*). DAOs are both *autonomous,* to the extent that they do not need (nor heed) their original creator, and *self-sufficient*, to the extent that they can charge users for their own services (or assets) in order to pay for the services they need.

DAOs (and smart contracts more generally) interact with the physical world through interfaces or sensors (so-called *Oracles*) that record information from the outside world into the blockchain. These are specifically relevant in the context of the Internet of Things, made of connected devices that constitute the interface between the physical and digital world. Any device connected to the Internet (or to a local network) can turn into "smart property" insofar as it can read the state of a blockchain and react to its changes over time (*e.g.* a "smart car" that only turns on if the driver possess a valid cryptographic token). With the emergence of blockchain-enabled devices, capable of interacting with one another, and with other smart contracts or DAOs on the blockchain, the Internet of Things might increase its potential effects on the physical world. This could lead to the emergence of complex ecosystems of smart devices, with humans and DAOs interacting with one another, often with unforeseeable consequences (Swan 2015).

## II. B Blockchain Code is Law

As every other technology, blockchain technology is not neutral. It is a technical artefact with a particular *architecture*, which inevitably has both social and political implications, as it facilitates certain actions and behaviors more than others. Besides, while blockchain technology presents a series of distinctive characteristics that differentiate it from other types of code, it enjoys nonetheless the same attributes of code (described above). The **barriers to entry** for smart contracts builders is low, which sets the ground for a potentially broad experimentation on an unexplored field. Just like any other piece of software, smart contracts feature a high degree of **malleability** and **adaptability**, enabling people to fork and experiment with a wide range of versions or adaptations of the same smart contracts. Blockchains are **transnational**, because they bypass the need for a central server (which necessarily needs to be located in a specific jurisdiction); and as smart contracts are deployed and executed on a distributed network of nodes, they significantly reduce the risk of prosecution or legal proceedings. Finally, smart contracts provide **ex-ante enforcement of technical rules**, thereby reinforcing the opportunities of regulation by code and the corresponding legal implications it might entail.

---

[9] Considering "organization" as an entity comprising multiple people (or SCs) with a specific goal, not a legally registered organization.

[10] Using cryptocurrencies such as Bitcoin significantly facilitates the transfer of currency by these pieces of software, i.e. smart contracts.

These implications are, however, to a large extent ignored by the current blockchain community. Discussion is currently very focused on the technical aspects of smart contract deployment and their implementation within a particular technological framework. Many smart contract proponents claim that, with the blockchain, many contractual clauses could be made partially or fully self-executing, self-enforcing, or both. The focus is mostly put on efficiency and optimization, with a view to provide a level of security that is superior to traditional contract law and to reduce other transaction costs associated with contracting.

To illustrate the extent to which blockchain code can assume the function of law, let us take the example of a hypothetical blockchain-based DRM system. Copyright law introduces "artificial scarcity" in the realm of information, by prohibiting (or constraining) the reproduction of creative works without the consent of the corresponding right holders. Yet, given the ease with which one can produce an identical copy of a digital work, copyright infringement has become widespread in the digital world. Since many years already, content providers have been relying on technological means (such as DRM systems, or other technological measures of protection) to restrain the way in which content can be accessed, used or reused by introducing a new set of technical rules, as a complement to the legal provisions of copyright law. Yet, most of these systems are limited by the fact that it is impossible to distinguish one digital file from another. By leveraging on the transparency and immutability of blockchain technologies, it is possible to restore the unicity and transferability of digital works, by linking every digital copy to a particular token on the blockchain. Authors can then associate these tokens with a particular set of rights to their digital works and trade them in the same way as they would trade digital tokens. Blockchain technology can thus be used to implement "artificial scarcity" at the level of each individual file—thus potentially allowing for the reintroduction of the first sale doctrine[11] in the digital realm, without the need to rely on any contractual or legal means.

Some proponents suggest that blockchain technology could lead to a society where self-enforcing rules would supplant traditional laws (Nakamoto, 2008). Indeed, with the advent of blockchain technology and the introduction of smart contract capabilities on top of it, it becomes increasingly appealing for people to bypass the traditional legal framework of contract law, and to rely on the underlying technical infrastructure provided by the blockchain instead. This was already the case for many DRM systems in the past, whose technical restrictions often extended beyond the scope of copyright protection, and this is likely to remain the case as more and more of our online interactions and economic transactions are mediated through smart contract code.

---

[11] The first sale doctrine (also known as the principle of exhaustion) stipulates that whoever has legitimately acquired a copy of a copyrighted work has the right to freely dispose of that copy, without seeking authorization from the copyright owner. The first sale doctrine has been enshrined into law both in the U.S. Copyright act (17 § 109) and in the Europe Directive 2001/29/EC on the harmonisation of certain aspects of copyright and related rights in the information society (Art. 4)

Yet, one important question that is often not accounted for within the blockchain community is whether smart contracts are in fact actionable in the real world. While they can be regarded, at their core, as a written contract drafted in a computer language, it is not clear —at this date— whether their code is "legally binding" upon the parties interacting with these contracts.

### II. C Law turning into Code

Regulation by code has acquired great momentum in the last few years because, as more and more of our interactions are mediated through technology, code has become much more efficient than law in its capacity to enforce rules. Hence, we are gradually delegating to technology the fundamental task of both interpreting and applying the law.

However, as we have seen with the case of DRM systems, it is not always easy to transpose legal rules (*wet code)* into technical rules (*dry code*). The former consists of an inherently ambiguous and flexible language, which can be applied on a case-by-case basis to an indefinite number of situations that might not have been precisely foreseen. The latter is made of a strict and formalized language, which requires well-defined categories and the precise stipulation of methods and conditions that need to be defined in advance. Despite the obvious discrepancies that subsist between these two typologies, it is becoming more and more common for legal rules to be translated into technical rules in order to be incorporated into a technological —hardware or software— device.

But as we increasingly rely on technological means to enforce legal rules, we face the risk that law progressively assume the characteristics of code, with rules becoming more and more formalized to better match the technology that is meant to enforce them.

With the advent of blockchain technology, this risk has become a reality —at least in the field of contracts. For a long time already, contractual provisions have been implemented directly into code (as in the case of traditional DRM systems) in order to facilitate the enforcement thereof. As time passes, and technology becomes a more and more attractive means to enforce contractual provisions, it might become increasingly less necessary to rely on actual legal contracts to support a variety of transactions. Besides, with the advent of smart contracts, code can be used not only for the purpose of *enforcing* existing legal provisions, but also with a view of *defining* them in the first place.

Indeed, blockchain-based DRM systems enable authors, artists and other copyright owners to enter into a direct relationship with the public, using smart contracts to establish the terms and conditions for accessing their works. When used in combination with blockchain-based payment systems, smart contracts make it possible for anyone to send micro-transactions to the relevant right holders in order to automatically obtain a license that will 'unlock' certain functionalities of the work (e.g. they might acquire the right to access, reproduce, or perhaps even remix a digital copy of the work), regardless of whether these functionalities are actually protected under the copyright regime. Smart contracts could also be deployed by collective right management societies in order to automate the collection of royalties to be paid to copyright owners whenever their works are performed, played or displayed in a public place. The distribution of copyright fees could thus be achieved in a much more transparent and efficient manner, with royalties being distributed to authors in real-time. What's more, the law could even introduce an obligation for certain actors to rely on smart contracts for internal logistics or accounting, with a view to automatize a variety of legal requirements. For instance, hardware and devices manufacturers might be required by the State to rely on a particular smart contract system in order to automatically redistribute the copyright levy[12] on private copying to the relevant right holders, without any third party intervention.

In view of that, and if it is true that on the cyberspace '*code is law*' (Lessig, 1999), we could say that with the advent of blockchain technologies, *law is progressively turning into code*.

The difference is a conceptual (rather than technical) one. Indeed, just as in the case of DRM systems, smart contracts can be regarded as the mere implementation of legal and technical rules into the code of a particular infrastructure or device. The trustless character of the blockchain has little to do with its ability to actually enforce these rules —except for the fact that it eliminates the need for a trusted intermediary to mediate any transaction. What makes the blockchain different from other technologies is that smart contracts are actually meant to replace legal contracts. They are no longer regarded as a mere support or enforcement mechanism to existing legal rules, rather, their code is intended to have the effect of law as its primary function.

Accordingly, as more and more contractual provisions are implemented in the form of a smart contract (as opposed to a legal contract), the blockchain progressively acquires the status of a "**regulatory technology**" —i.e. a technology that can be used both to *define* and *incorporate* legal or contractual provisions into code, and to *enforce* them irrespectively of whether or not there subsists an underlying legal rule.

There are, of course, many important issues that should be accounted for in the process of rethinking law through the lenses of technology.

First of all, laws cannot (or should not) be entirely and exclusively defined through technological processes, as technology cannot replace the democratic debate which necessarily takes place within the legislative branch. While the legal system implements a series of policies and procedures for society to collectively agree upon the type of rights and obligations that people ought to comply with, technical rules can be unilaterally imposed by software builders and device producers, without following any democratic debate.

---

[12] The copyright levy on private copying (also known as the "blank media tax") is a government-mandated tax that is charged (in addition to the TVA) on the purchases of recordable media, such as CDs, DVDs, USB keys or hard-drives.

Besides, the legal system ensures the existence of a public, transparent and explicit set of *universal* rules, whose legitimacy can easily be put into question. In contrast, regulation by code is elaborated mostly by private actors, who incorporate a set of *arbitrary* rules into technical artefacts, without any public purview and often without giving the opportunity for people to put these rules into question (this is especially true in the case of proprietary software that does not publish its source code).

Secondly, while it is true that contract law allows for multiple parties to enter into an agreement to establish a voluntary set of rules that represent the will of these parties, contract law also incorporates a series of legal safeguards that might either invalidate the contract or make it non-enforceable, as a result of failure to comply with specific formalities (e.g. mutual consent, consideration, capacity, legality) or due to a vice in the contract (as a result of e.g. unconscionability, undue influence, or coercion). These legal safeguards are generally not incorporated directly into the wordings of the contract, but they are nonetheless accounted for through the judicial system. In the context of smart contracts, since the enforcement is done through the technological framework itself, it becomes possible for private parties to bypass these legal safeguards (just like DRM systems commonly bypass copyright *fair use* provisions). Any smart contract that is technologically sound will be enforced, regardless of whether or not it qualifies as a valid contract under the law.

Moreover, while specific software tools can support the drafting of better legislation (*e.g.* expert systems designed to identify conflicts or fallacies within the legal system), technology can also bring us to re-think the way in which the law is being written, encouraging a shift towards a more quantitative and/or formalized approach to law drafting.

This modern approach to lawmaking presents a few benefits, especially as it reduces the ambiguities (and therefore the gray areas) characteristics of traditional legal drafting. In the long run, the lack of textual ambiguity might reduce the need for canons of construction and other textual interpretation techniques —although factual ambiguity (*i.e*. did a real world event happen or not) will obviously remain.

Indeed, one must not forget that blockchain-based applications are meant to operate in the "real world" —one that is regulated by traditional rules of law. While smart contracts are potentially able to handle complex deal logics, many kinds of transactions do eventually have to interface with people or organizations that subsist in the physical world. This is problematic to the extent that it can reduce (or even eliminate) the *trustlessness* of the transaction. For instance, even if it is possible to transfer property titles via blockchain transactions (e.g. by transferring the *cryptographic title* to a smart car), the blockchain alone is unable to ascertain whether the property has actually been transferred in the real world (e.g. whether the car has been *physically* and *legally* transferred to the new owner), or whether it was perhaps faulty or defective, etc. While it is, of course, possible to implement an articulate system of collaterals, which might considerably increase the complexity of these contracts, smart contracts will always and necessarily have to rely on a trusted intermediary (or *'oracle'*) whenever they need to interface with the real world to provide external validation. It is at those "choke points" that the legal system ultimately have a say in the context of a breach. In order to be as effective as their traditional counterparts, smart contracts must in fact also be actionable in the real world. This might, of course, require people to comply with all the standard formalities required for a court to enforce a contract under the applicable law.

Finally —and perhaps most importantly— it is important to understand the consequences of having legal and contractual provisions being drafted or elaborated *as code*, as opposed to simply being implemented *into code*.

Many legal rules are intended to be generic enough to be applicable to a variety of different situations —some of which could not have been foreseen at the time of drafting these rules. Indeed, in order to be sustainable over time, legal rules need to be highly generic. They must be drafted at a higher layer of abstraction so as to be agnostic to the specificities of a case. They must be generic enough to be able to encompass new and unforeseen situations, which are factually different from previous cases but which are practically or ideologically the same.

Of course, the more generic a rule is, the easier it will apply to edge-cases —but, at the same time, the more likely it will also apply to cases which were actually not meant to be covered by that rule. This is the reason why legal rules generally need to be interpreted and construed by a judge, before they can be applied, on a case-by-case basis, to the facts of a case.

For a long time, law has been drafted *by* humans and *for* humans. Human judgement is thus necessary in order to give meaning to the law. In particular, in order to properly appreciate the wording of the law, it is essential to account for the original intentions of the legislator — something that requires a general understanding of the context and contingencies that existed at the time in which the law was drafted.

Given this inherent ambiguity and flexibility, the automatic execution of legal and contractual provisions cannot be achieved without formalizing these rules into a more formal language which can be processed and understood by a machine. Yet, turning natural language into a formal language is a complicated task that will almost unavoidably limit the scope of application of these rules, given that the translation entails a necessary loss in generality.

In order to facilitate this process, we observe, in recent years, a significant transformation in the practices of legal drafting. Both legal provisions and contractual clauses are gaining in specificity, their wording is becoming increasingly precise, and their interpretation is consequently becoming much more objective than before. These clauses are therefore much easier to incorporate into code, so as to be automatically enforced by technological means.

Yet, this drive towards an increasingly formalized language goes against the traditional conception of law, perceived as an inherently flexible and ambiguous set of rules. Although the judicial system is subject to the prerequisites of neutrality and impartiality, the quest for an objective rule of law has often been criticized (MacCormick, 2009) on the grounds that the meaning of the law must always (and necessarily) be construed according to the fact of the case, following the interpretation of a judge.

**Conclusion**

In the last decades, regulation through code has become an increasingly popular tool to regulate the behavior of people, both online and offline. With the growing centrality that digital technologies have acquired in our everyday lives, code is now capable of regulating and constraining our actions in a wide variety of ways. On the Internet, in particular, code has been repeatedly used to implement different sets of affordances and constraints (Benkler, 2006), and to incorporate certain values into code in ways that can deeply affects us (see *e.g.* the case of P2P file sharing systems, enabling people to cooperate and share information with one another; or the case of Facebook, promoting individualistic and narcissistic values, combined with a culture of control and surveillance). This form of regulation is frequently happening without any awareness from users affected by it.

The advent of blockchain technologies is a notable step towards the widespread adoption of technical regulation. The new possibilities provided by blockchain technology provide a new field for experimentation and innovation, which has been framed by some as a new "hype" (Reber and Feuerstein, 2014). Yet, the blockchain is regarded by many financial actors as an ideal technology for optimizing a variety of existing financial applications and for enabling new types of financial technology (*fintech*) services. The blockchain is also considered as a useful technology in the realm of Internet of Things insofar as it can provide a common playing field through which connected devices can easily interact (and transact) with one another (Hajdarbegovic, 2014).

There are, indeed, exciting potentials to explore with this technology —yet, there are just as many scary scenarios we may easily step in. With the blockchain, the legal challenges which were raised in the past by DRM systems could now be raised once again (with a much broader impact this time) by blockchain-enabled devices, operating according to technical rules dictated by smart-contracts on a blockchain (e.g. door locks opening only when presented with a valid cryptographic token, self-driving cars negotiating speed on the highway, etc.). Given the *ex-ante* enforcement of regulation by code, combined with the lack of flexibility of its technical rules, blockchain-enabled devices cannot distinguish between routine situations and edge-cases that might require a different type of treatment (e.g. the need for opening a door in the event of a fire, or speeding up to rescue a wounded person).

Law is intentionally ambiguous, so that it can be more easily applied on a case-by-case basis. It is the overlapping of multiple legal provisions which creates a solid regulatory framework, with multiple limitations and exceptions in order to accommodate the complexity and unpredictability of human society.

Conversely, code is extremely strict and intrusive in its enforcement mechanisms. Hence, if not properly designed, regulation by code might actually oppose the interest of the individual it is meant to regulate.

Thus far, law has found ways to regulate the code in order to limit its disruptive potential. However, the decentralized nature of the blockchain and the resulting attributes of smart contract code (which can be used to create autonomous, self-sufficient, and potentially unstoppable DAOs) raise new important issues in terms of legal responsibility and regulability. Just as the law cannot prevent a biological virus from spreading, it also cannot shutdown autonomous software agents simply by ordering them to do so.

The prospect of automated legal governance is something that should, to the very least, be examined with great caution, as it might be opening novel scenarios, whose consequences simply cannot foresee. Most importantly, by automating the enforcement of the law, we may perhaps gain in efficiency and transparency, but we might eventually also reduce the freedoms and autonomy of individuals (De Filippi and Wright, 2015). As Lessig (2006) has elegantly put it:

> *"[The code] will present the greatest threat to both liberal and libertarian ideals, as well as their greatest promise. We can build, or architect, or code cyberspace to protect values that we believe are fundamental. Or we can build, or architect, or code cyberspace to allow those values to disappear. There is no middle ground. There is no choice that does not include some kind of building. Code is never found; it is only ever made, and only ever made by us." (Lessig, 2006)*

Accordingly, there are important tradeoffs to be accounted for in the process of turning law into code. While allowing for anyone to implement and deploy their own techno-legal frameworks has strong democratic potential, if coopted by the current economic or political order, the process might possibly lead to a regime of inflexible (perhaps even totalitarian) networked governmentality. This controversial scenario may materialize itself as an utopia or a (crypto-)libertarian dream, but it might also lead to a dystopian society featuring a strong and decentralized panopticon.

**About the authors**
Primavera De Filippi (PhD) is a permanent researcher at the National Center of Scientific Research (CNRS) in Paris. She is faculty associate at the Berkman Center for Internet & Society at Harvard Law School, where she is investigating the concept of "governance-by-design" as it relates to online distributed architectures. Most of her research focuses on the legal challenges raised, and faced by emergent decentralized technologies —such as Bitcoin, Ethereum and other blockchain-based applications —and how these technologies could be used to design new governance models capable of supporting large-scale decentralized collaboration and more participatory decision-making. Primavera holds a PhD from the European University Institute in Florence. She is a member of the Global Future Council on Blockchain Technologies at the World Economic Forum, as well as the founder of the Internet Governance Forum's dynamic coalitions on Network Neutrality, Platform Responsibility and Blockchain Technology (COALA). In addition to her academic research, Primavera acts as a legal expert for Creative Commons in France and sits on the stakeholder board of the P2P Foundation.
Follow Primavera on Twitter: @yaoeo

Samer Hassan (PhD) is an activist and researcher, Fellow at the Berkman Center for Internet & Society (Harvard University) and Associate Professor at the Universidad Complutense de Madrid (Spain). Currently focused on decentralized collaboration, he has carried out research in decentralized systems, social simulation and artificial intelligence from positions in the University of Surrey (UK) and the American University of Science & Technology (Lebanon). Coming from a multidisciplinary background in Computer Science and Social Sciences, he has more than 45 publications in those fields (H-index=11). Engaged in free/open source projects, he co-founded the Comunes Nonprofit and the Move Commons webtool project. He's an accredited grassroots facilitator and has experience in multiple communities and grassroots initiatives. He's involved as UCM Principal Investigator in the EU-funded P2Pvalue project on building decentralized web-tools for collaborative communities and social movements. Follow Samer on Twitter: @samerP2P
E-mail: samer [at] fdi [dot] ucm [dot] es

**Acknowledgements**
This work was partially supported by the Framework programme FP7-ICT-2013-10 of the European Commission through project P2Pvalue (grant no.: 610961), and by the Spanish Ministry of Education's Jose Castillejo program for researcher mobility 2015 (ref. CAS15/00203).

**References**
1. Philippe Aigrain, 2012. Sharing: Culture and the economy in the internet age. Amsterdam University Press.
2. John Perry Barlow, 1996. A Declaration of the Independence of Cyberspace.
3. Yochai Benkler, 2006. *The wealth of networks: How social production transforms markets and freedom.* Yale University Press.
4. Yochai Benkler., Hal Roberts, Rob Faris, Alicia Solow-Niederman, and B. Etling, 2015. Social mobilization and the networked public sphere: Mapping the SOPA-PIPA debate. Political Communication, volume 32, number 4, pp. 594-624.
5. Robert C. Berring, 1986. Full-text databases and legal research: Backing into the future. High Technology Law Journal, volume 1, number 1, pp. 27-60.

6. June M. Besek, 2003. Anti-Circumvention Laws and Copyright: A Report from the Kernochan Center for Law, Media and the Arts. Colum. JL & Arts, volume 27, number 385.
7. Vitalik Buterin, 2014. Ethereum: A next-generation smart contract and decentralized application platform. Whitepaper.
8. Martin Campbell-Kelly, 1995. Development and Structure of the International Software Industry, 1950-1990. Business and Economic History, volume 24, number 2, pp. 73–110.
9. Primavera De Filippi and Aaron Wright, 2015. Decentralized blockchain technology and the rise of lex cryptographia. Available at SSRN 2580664.
10. Kristin R. Eschenfelder, and Anuj C. Desai, 2004. Software as protest: The unexpected resiliency of US-based DeCSS posting and linking. The Information Society, volume 20, number 2, pp. 101-116.
11. Charles L. Evans, 1993. US Export Control of Encryption Software: Efforts to Protect National Security Threaten the US Software Industry's Ability to Compete in Foreign Markets. NCJ Int'l L. & Com. Reg., volume 19, number 469.
12. Mayo Fuster Morell, 2012. The free culture and 15M movements in Spain: Composition, social networks and synergies. Social Movement Studies, volume 11, number 3-4, pp. 386-392.
13. Tarleton Gillespie, 2014. The Relevance of Algorithms. Media technologies: Essays on communication, materiality, and society, volume 167.
14. Nermin Hajdarbegovic, 2014. Ibm sees role for block chain in internet of things. Coindesk Magazine, September 2014.
15. Lucas Introna and Niall Hayes, 2004. Plagiarism, detection and intentionality: on the construction of plagiarists. Plagiarism: Prevention, Practice and Policies, 28th June-30th June.
16. Bernward Jeorges, 1999. Do Politics Have Artifacts?. *Social Studies of Science*, 29, 411-31
17. Gopinaath Kannabiran, and Marianne Graves Petersen, 2010. Politics at the interface: A Foucauldian power analysis. Proceedings of the 6th Nordic Conference on Human-Computer Interaction: Extending Boundaries, pp. 695-698. ACM.
18. Adam D. Kramer, Jamie E. Guillory, and Jeffrey T. Hancock, 2014. Experimental evidence of massive-scale emotional contagion through social networks. *Proceedings of the National Academy of Sciences*, volume *111*, number 24, pp. 8788-8790.
19. Raymond Shim Ray Ku, 2002. The creative destruction of copyright: Napster and the new economics of digital technology. *The University of Chicago Law Review*, pp. 263-324.
20. Eszter Hargittai, 2010. Facebook privacy settings: Who cares?. First Monday, volume 15, number 8.

21. Lawrence Lessig, 1997. Constitution of Code: Limitations on Choice-Based Critiques of Cyberspace Regulation, The. CommLaw Conspectus, volume 5, number 181.
22. Lawrence Lessig, 1999. *Code and other laws of cyberspace*, volume 3. New York: Basic books.
23. Lawrence Lessig, 2006. Code. New York: Basic books.
24. Geert Lovink and Miriam Rasch, 2013. *Unlike us reader: Social media monopolies and their alternatives* No. 8. Institute of Network Cultures.
25. Neil MacCormick, 2009. *Rhetoric and the rule of law: a theory of legal reasoning.* OUP Oxford.
26. Duncan Matthews and Petra Žikovská, 2013. The rise and fall of the anti-counterfeiting trade agreement (ACTA): lessons for the European Union. IIC-International Review of Intellectual Property and Competition Law, volume 44, number 6, pp. 626-655.
27. Abbe Mowshowitz, 1984. Computers and the myth of neutrality. *Proceedings of the ACM 12th annual computer science conference on SIGCSE symposium*, pp. 85-92. ACM.
28. Satoshi Nakamoto, 2008. Bitcoin: A peer-to-peer electronic cash system.
29. Daniel Neyland, 2006. Privacy, surveillance and public trust. New York: Palgrave Macmillan.
30. Frank Pasquale, 2015. *The black box society: The secret algorithms that control money and information*. Harvard University Press.
31. Andrew Patrick and Steve Kenny, 2003. From privacy legislation to interface design: Implementing information privacy in human-computer interactions. Privacy Enhancing Technologies, pp. 107-124. Berlin/Heidelberg: Springer.
32. Johan Pouwelse, Pawel Garbacki, Dick Epema and Henk Sips, 2005. The bittorrent p2p file-sharing system: Measurements and analysis. *Peer-to-Peer Systems IV,* pp. 205-216. Springer Berlin Heidelberg.
33. Daniel Reber and Simon Feuerstein, 2014. Bitcoins-Hype or Real Alternative?. *Internet Economics VIII*, volume 81.
34. Joel R. Reidenberg, 1997. Lex informatica: The formulation of information policy rules through technology. *Tex. L. Rev.*, volume *76*, number 553.
35. Sam Ricketson, 1987. Berne Convention for the Protection of Literary and Artistic Works: 1886-1986. Centre for Commercial Law Studies, Queen Mary College: Kluwer.
36. Jeremy Rifkin, 2014. The zero marginal cost society: The internet of things, the collaborative commons, and the eclipse of capitalism. Macmillan.
37. Bill Rosenblatt, Stephen Mooney and Bill Trippe, 2001. Digital rights management: business and technology. John Wiley & Sons, Inc.
38. Pamela Samuelson, 2003. DRM {and, or, vs.} the law. *Communications of the ACM*, volume *46*, number 4, pp. 41-45.

39. Neil Smith, 1996. The new urban frontier: gentrification and the revanchist city. Psychology Press.
40. Melanie Swan, 2015. *Blockchain: Blueprint for a New Economy*. 'Reilly Media, Inc.
41. Nick Szabo, 1997. Formalizing and securing relationships on public networks. *First Monday*, volume *2*, number 9.
42. Donald A. Waterman, Ray J. Paul and Richard M. Peterson, 1986. Expert systems for legal decision making. Expert Systems, volume 3, number 4, pp. 212-226.
43. Langdon Winner, 1980. Do artifacts have politics?. *Daedalus*, pp. 121-136.
44. Steve Woolgar and Robert G. Cooper, 1999. Do artefacts have ambivalence? Moses' bridges, Winner's bridges and other urban legends in S&TS. Social studies of science, volume 29, number 3, pp. 433-449.
45. Tim Wu, 2003. When code isn't law. Virginia Law Review, pp. 679-751.
46. Malte Ziewitz, 2016. Governing Algorithms Myth, Mess, and Methods. Science, Technology & Human Values, volume 41, number 1, pp. 3-16.